# Amplifying Evanescent Waves by Dispersion-induced Plasmons: Defying the Materials Limitation of Superlens

Tie-Jun Huang, Li-Zheng Yin, Jin Zhao, Chao-Hai Du, Pu-Kun Liu[*]

State Key Laboratory of Advanced Optical Communication Systems and Networks, Department of Electronics, Peking University, Beijing, 100871, China

*Correspondence to pkliu@pku.edu.cn

**Breaking the diffraction limit is always an appealing topic due to the urge for a better imaging resolution in almost all areas. As an effective solution, the superlens based on the plasmonic effect can resonantly amplify evanescent waves, and achieve subwavelength resolution. However, the natural plasmonic materials, within their limited choices, usually have inherit high losses and are only available from the infrared to visible wavelengths. In this work, we theoretically and experimentally demonstrate that the arbitrary materials, even air, can be used to enhance evanescent waves and build low loss superlens with at the desired frequency. The operating mechanisms reside in the dispersion-induced effective plasmons in a bounded waveguide structure. Based on this, we construct the hyperbolic metamaterials and experimentally verified its validity in the microwave range by the directional propagation and imaging with a resolution of 0.087λ. We also demonstrate that the imaging potential can be extended to terahertz and infrared bands. The proposed method not only break the conventional barriers of plasmon-based lenses, but also bring possibilities in applications based on the enhancing evanescent waves from microwave to infrard wavelengths, such as ultrasensitive optics, spontaneous emission, light beam steering.**

## Introduction

Imaging with the unlimited resolution has been an intriguing dream of scientists for a few centuries. The diffraction limit indicates that the features smaller than half of the operating wavelength is carried by evanescent waves which are permanently lost in conventional imaging processes [1]. With the assist of an elaborate probe, the near-field scanning microscopy can capture the lost information and reach nanometer resolution [2-4]. Although this imaging technique has been highly commercialized, it still faces the time-consuming problem and is not suitable for many applications. In 2000, the plasmonic effect was demonstrated as an efficient method for breaking the diffraction limit [5], due to its ability of resonantly amplifying evanescent components [6]. After that, the near-field superlenses are experimentally demonstrated from far-infrared to optical wavelengths by metal slabs, graphene and other dielectrics [7-12]. In further investigations, by stacking plasmonic materials and dielectrics, the anisotropic plasmonic materials, also called hyperbolic metamaterials (HMM), was also developed for superresolution imaging [13-17]. HMM can provide lower propagating loss and larger transmission band of spatial spectrum because of the continuously exciting of surface-plasmon polaritons (SPPs) at metal-dielectric interfaces [18, 19].

Clearly, the plasmonic effect is essential for beating the diffraction limit. However, within their limited choices, natural plasmonic materials often suffer from intrinsic high loss [1, 20]. Besides, when the frequency comes to microwave and terahertz regimes, plasmonic effects of most materials become negligible and the metals behave as perfect electric conductors in this range [21-23]. To develop the low-loss and flexible plasmons at low frequencies, the concept of spoof surface plasmons polaritons (SSPPs) [24] and effective surface plasmons polaritons (ESPPs) [25], are respectively developed by periodically texturing the geometry of metals and tailoring the structural dispersion of electromagnetic (EM) modes in bounded waveguides. These methodologies have provided enormously possibilities in sensing [26, 27], focusing [28, 29], guiding [30-32], on-chip sources [33], generating vortex beams [34, 35], and novel applications based on controlling waves in subwavelength scale [36-39]. Recently, we also proposed the method of spatial spectrum sampling by SSPPs [40], a method for retrieving the broadband evanescent information of targets and realizing subwavelength resolution imaging. Here, the SSPPs structure was employed as a sensitive spatial filter. Nevertheless, for imaging applications the designed plasmons are mainly employed as a probe, whether for directly detecting or extracting spatial spectrum, which is inconvenient for the non-destructive testing and biomedicine diagnosis at microwave and terahertz wavelengths. Amplifying broadband evanescent waves, the key factor of building superlens and many other exotic phenomena, is still a blank for the designed plasmons so far, to the best of our knowledge.

In this work, we made full use of the ESPPs and its resonance properties to break the diffraction limit. It showed that by tailoring the structural-induced dispersion of foundational TE mode in a parallel-plane waveguide (PPWG), the evanescent waves could be efficiently amplified by arbitrary dielectrics, even air. Based on this, an 'air' superlens with resolution exceeding $0.1\lambda$ was demonstrated. As a specified realization, by a periodic stack of air and common dielectric layers, we also constructed the effective HMM and fully studied its potentials. The HMM was numerically verified from microwave to infrared wavelengts by two phenomena, the directional propagation and the imaging with a calculated resolution of $0.04\lambda$. The experiments performed in the microwave regime matched well with the theoretical predictions, and a measured resolution of $0.086\lambda$ was obtained. This work showed that the low-loss superlensing effect could be practically realizable at any frequencies by available dielectric materials.

## Results and Discussion

**Amplifying the Evanescent Waves by ESSPs**

Firstly, we briefly study the properties of ESPPs supported by a PPWG, as shown in Fig. 1(a). It is filled with two isotropic dielectrics, with real permittivities being $\varepsilon_{r1}$ and $\varepsilon_{r2}$ ($\varepsilon_{r1}, \varepsilon_{r2} > 0$). Normally, the SPPs are supported on the interface of two materials with opposite values of permittivity. This demand can be met through geometric dispersion of conventional dielectric inside the PPWG. The propagation of TE1 mode along the PPWG leads to geometric dispersion of the dielectrics, the effective permittivity $\varepsilon_e$ of which becomes

$$\varepsilon_e = \varepsilon_r - \frac{\lambda^2}{4a^2}, \tag{1}$$

where $\lambda$ is the operating wavelength and $a$ is the separation between two parallel plates. The metallic wires with period $p$ placed in the interface between two materials can prevent the generation of TM (transverse magnetic) modes and accumulate electric charges for the normal components of the electric field. Theoretically, the filled materials are arbitrary. In this work, the dielectric 1 is set as most common dielectric, air, which seems impossible to amplify the evanescent waves. For simplicity, the calculated frequency and the radius of the wires are kept as 2.5 GHz and $\lambda/300$. The commercial FEM (finite element method) software COMSOL Multiphysics is employed to calculate the EM field evolution. In the first case, the parameters are set as $p = \lambda/40$, $a = 34.6$ mm, $\varepsilon_{r2} = 4.3$. Then, the effective permittivities of the chosen materials turn out to be $\varepsilon_{e1} = -2$, $\varepsilon_{e2} = 1.3$. The field distributions of $E_x$, $E_y$ and $H_z$ at the middle plane of the PPWG are illustrated in Fig. 1(b), respectively related to b1-b3. The EM energy shows high confinement and propagates in the form of surface waves at the interface, well consistent with natural SPPs. The simulated wavelength of ESPPs is 0.525$\lambda$, also matched with the theoretical calculation 0.519$\lambda$ by $\lambda_{spp} = \lambda\sqrt{(\varepsilon_{r1} + \varepsilon_{r2})/\varepsilon_{r1}\varepsilon_{r1}}$. In the b4 of Fig. 1(b), the snapshot of $H_y$ is rendered, the sampled y-z plane of which is located at 0.4 mm from the cylinders. One can see that the $H_y$ component has non-zero intensity when it leaves the middle x-y plane of the PPWG, which is quite different from traditional SPPs. Actually, affecting by the PPWG, the complex propagating mode has five components [38], which are $H_x$, $H_y$, $H_z$, $E_x$ and $E_y$. In the middle plane, $H_x$ and $H_y$ componts will vanish, indicating that the EM waves degenerate into the TM-polarized waves rather than the original TE waves. Therefore, the effective permittivity and the phenomenon of ESPPs are only valid at the middle plane of the PPWG. In the following studies, we only investigate the physical mechanisms on this plane.

To study the behavior of amplifying evanescent waves, a structure is designed and rendered in Fig. 2(a), where the middle air and dielectric at both sides are used for mimicking the plasmonic and positive (refers to the value of permittivity) materials. When a TM-polarized wave with parallel wavevector $k_x$ impinges on a homogenous slab, the transmission coefficient $T$ is expressed by (see section 1 in supplement).

$$T = \frac{2}{2\cos k_{y2}d - i\left(\frac{\varepsilon_{r1}k_{y2}}{\varepsilon_{r2}k_{y1}} + \frac{\varepsilon_{r2}k_{y1}}{\varepsilon_{r1}k_{y2}}\right)\sin k_{y2}d}, \quad (2)$$

where $k_{y1} = \sqrt{\varepsilon_{r1}k_0^2 - k_x^2}$, $k_{y2} = \sqrt{\varepsilon_{r2}k_0^2 - k_x^2}$, $k_0 = 2\pi/\lambda$, $d$ is the thickness of the slab, $\varepsilon_{r2}$ and $\varepsilon_{r1}$ are the permittivities of this slab and the surrounding material. Clearly, when $\varepsilon_{r2} = -\varepsilon_{r1}$, the enhanced transmission with broadband $k_x$ can be obtained, resulting from the significant surface charges accumulation at the plasmon-dielectric interface. In this situation, as indicated in the insert of Fig. 2(a), the evanescent waves will be amplified at two interfaces, which is essential to break the diffraction limit. To confirm if the ESPP also has this capability, we choose the parameters $a = 34.6$ mm, $\varepsilon_{r2} = 5$, and $d = 14$ mm to perform the simulation, corresponding to $\varepsilon_{e1} = -2$, $\varepsilon_{e2} = 2$. A 0.4 mm-width slit is employed as the source. The distance between the source and the recording plane is $2d$ and the air slab is located in the center of two planes. Fig. 2(b) presents the calculated electric field distribution ($|E|$) of effective plasmonic materials. By exciting the ESSP, significant field enhancement at both interfaces is observed, matched well with real plasmonic materials (Fig. 2(c)). This enhancement gives direct evidence for the amplification of evanescent components. To quantitatively show the enhanced transmission, we define an optical transfer function $T_h$, which is the ratio of the spatial harmonics in the source plane to the recording plane, $|H_z(k_x)_{rec}|/|H_z(k_x)_{sou}|$. The intensity of spatial harmonics is obtained by Fourier transform of the simulated fields in Fig. 2(b). In Fig. 2(d), we plot the $|T_h|$ versus parallel wavevector from 0 to $7k_0$ (dashed line).

Clearly, by exciting ESPP at the resonance condition $\varepsilon_{e1} = -\varepsilon_{e2}$, the broadband amplification of evanescent waves can be realized. The trend of $T_h$ is also consistent with the theoretical calculation (see section 1 of the supplement).

Here, we want to emphasize that our work is totally different from previous studies of dielectric superlenses [9-11, 41]. In an attempt to the realistic plasmonic applications, new constituent materials with low resistive loss have been extensively explored for decades. The promising no-metal materials including graphene, transparent conducting oxides [42-44], silicon carbide [10], gallium arsenide [45], boron nitride [46], perovskite [41] and other high-doping semiconductors [47, 48] have already shown extraordinary prospects, such as low loss, flexible tunability, low fabrication demand and high integration. These works mainly aim to demonstrate the intriguing properties a specified material, the working range of which is usually limited between the infrared to visible wavelengths. Note that, in this paper, we propose a method to break the material and wavelength limitations for the superresolution imaging area. The 'air' lens in the microwave frequency range is a simple realization of our method. In the discussion part, we will extend our method to the terahertz and infrared regimes with other dielectrics.

**Build HMM by ESSPs**

To further demonstrate the advantages of our mechanisms, in this section, we will build an effective HMM, a more practical and specific model. Our designed structure is illustrated in Fig. 3(a), consisting of periodic air and dielectric layers (each layer holds $\lambda/60$ thickness) in y direction. The wires with $\lambda/50$ period are arranged at all the interfaces. The whole system is treated as a homogeneous medium with anisotropic parameters $\varepsilon_x = (\varepsilon_{e1}+\eta\varepsilon_{e2})/(1+\eta)$ and $\varepsilon_y = (1+\eta)\varepsilon_{e1}\varepsilon_{e2}/(\varepsilon_{e1}+\eta\varepsilon_{e2})$. $\eta$ is the thickness ratio of the two layers. By adjusting the parameters, we can achieve $\varepsilon_x > 0$ and $\varepsilon_y < 0$. According to the dispersion relation $k_x^2/\varepsilon_y + k_y^2/\varepsilon_x = k_0^2$, a hyperbolic-shaped dispersive line is obtained. This indicates $k_y$ is real for an arbitrary large value of $k_x$. This indicates that the spatial harmonics even with very larger $k_x$, which are normally evanescent, are also supported inside in the HMM. The physical mechanism is that SSPs are excited at the interface of each layer, resulting in continuously amplifying the spatial components. Firstly, the well-known directional propagation of HMM is applied to demonstrate the validity of our structure. The parameters are set as: $a$ = 34.7 mm, $\varepsilon_{r2}$ = 9, leading to $\varepsilon_x$ = 2 and $\varepsilon_y$ = -6. The related dispersive curve (black solid line) is plotted in Fig. 3(c). Since the group velocity direction is perpendicular to the dispersive curve, the spatial harmonics with high $k_x$ inside HMM only have two preferred directions, as indicated by the red arrows. Clearly, the in-coupled beam will propagate in the highly directional form, rather than omnidirectional diverging. The directional propagation has been applied to nanolithography [18], superfocusing [19] and super-resolution imaging [49]. This phenomenon is also visualized by our effective HMM, as shown by the simulated field map in Fig. 3(b), where the EM waves generated by a $\lambda$/20-length slit is split into two directions. Therefore, the ESSP-based structure is capable of mimicking the directional propagation behavior. However, it is found that the beams in Fig. 3(b) slightly diverge when they go through multilayers, and the angle between the two beams is smaller than the theoretical value $\theta = \arctan\left(\sqrt{-\varepsilon_y/\varepsilon_x}\right)$. To calculate the dispersion of simulated results, we apply a two-dimensional Fourier Transform and depict the result in Fig. 3(c) by pseudo-color image. It is clear to see that the constructed materials allow the propagation of spatial harmonics with broadband $k_x$. However, with the increase of $k_x$, the response to our materials deviates from the ideal hyperbolic dispersion. The effective medium theory is inaccurate because the thickness of each layer is not much smaller than the working wavelength. Therefore, a new dispersion equation based on transfer matrix is employed to describe this structure

$$\cos(k_z(d_1+d_2)) = \cos(k_{z1}d_1)\cos(k_{z2}d_2) - \frac{1}{2}\left(\frac{k_{z1}}{\varepsilon_{e2}} + \frac{k_{z2}}{\varepsilon_{e1}}\right)\sin(k_{z1}d_1)\sin(k_{z2}d_2), \quad (2)$$

where $d_1$ and $d_2$ are the thickness of two dielectrics. More details about Equation 3 and its relation with effective medium theory is put in section 2 of supplement. In Fig. 3(c), we also plot the dispersive curve using the revised values (dashed line) which is well matched with the pseudo-color image.

Allowing propagating broadband harmonics with low-loss makes HMM become a better candidate for superlens. In this case, the dispersion of HMM should be as flat as possible, so that the harmonics with difference $k_x$ will propagate with the same group velocity away from the sources. The flat dispersion is realized by setting $\varepsilon_{r2} = 5.05$, and the corresponding dispersion relation is plotted in Fig. 4(a), where the group velocity for all waves is directed into the y direction. Two slits with $\lambda/50$ width and $\lambda/5$ center-to-center distance are employed as the imaging targets. With considering the disturbance introduced by loss issues, $\varepsilon_{r2}$ is reset as $5.05+0.1i$, the value of which exceeds the loss of most of the dielectrics in microwave frequency range. The calculated field snapshot of $H_z$ is depicted in the Fig. 4(b). It is clear to see that EM energy propagates along the y direction without x-direction spread, leading to the exact reproduction and transfer of input patterns. This behavior is well-known canalization imaging. Although the amplitude gradually decreases, two targets are totally separated in the $\lambda/2$-away imaging plane. Its ultimate resolution can be estimated by full-width at half-maximum of one peak in the 1-D autocorrelations of field. In Fig. 4(c), we illustrate the field intensity of the targets in the imaging plane. Then, the simulated ultimate resolution is about $0.04\lambda$. The resolution can be further improved by choosing low-loss materials and optimizing the geometrical parameters. The more practical far-field superresolution imaging is also verified by setting the lens into a cylindrical geometry (see section 3 of supplement).

The directional propagation and superresolution imaging fully demonstrate that the HMM can be induced by the ESPP even at microwave wavelengths. As the HMM has been applied in tremendous topics [18, 50] as subwavelength imaging, spontaneous emission enhancement, heat transport, on-chip sources, wave manipulation and so on, introducing new methods to construct HMM away from its original optical wavelength can open tremendous potentials. In addition, the HMM induced by ESPP has great advantages over the natural HMM for the flexible design and negligible metal loss at the calculated frequency. For example, the effective HMM can be realized by one dielectric, as depicted in Fig. 4(d). The different permittivity values can be realized by utilizing the PPWG with different thicknesses.

**Experimental Verification**

In this parts, we experimentally verify the exotic properties of the ESPP. Here, we mainly focus on the demonstration of effective HMM. Assembling the cylinder array and dielectric layers in three dimensions with high accuracy is very challenging, as the size of these objects is located in the deep-subwavelength domain. As shown in Fig. 5(a), we alleviate this dilemma by replacing the cylinders with ultra-thin metallic stripes, which is precisely integrated with dielectric layers by standard printed circuit board fabrication process. The ultra-thin stripes also have great advantages in developing conformal plasmonic devices. The width, thickness and height of the dielectric substrate are 120 mm, 36.7 mm and 2 mm ($\lambda/60$), respectively. To match the simulated conditions, the substrate is made by composite materials, whose permittivity can be designed from 3 to 22 by changing the filling ratio. In this work, we choose two samples with permittivity being 4.7 and 8.5, which are respectively corresponding to the simulations in Fig. 3(b) and Fig. 4(b). Note that, the electrolytic copper-clad stripes with 35 um thickness are located on both sides of the dielectric. A close shot of the stripes is illustrated in the insert, indicating that the width and period of the stripes are 0.5 mm and 2.4 mm. The copper plate, as shown in Fig. 5(b), is employed to construct the PPWG. The grooves on the plate with 2 mm width, 4 mm period and 1 mm height, are cut by lathe to fix the dielectric layers. The width of the plate is 66 mm, corresponding to 16 periods of grooves. It should be pointed out that, when calculating the effective permittivity, the value of $a$ for air should be 2 mm less than dielectric layers owing to the influences of grooves. Finally, by combining all parts, the picture of constructed multilayers structure is shown in Fig. 5(c).

The experimental system consists of a four-port millimeter wave vector network analyzer, two monopole antennas, a two-dimensional electric-controlled stage and its controller, fabricated HMM, absorber, and a computer, as shown in Fig. 5(d). In the measurement, two monopole antennas are employed as the emitting source and recording probe. The EM waves generated by the source will have interaction with the testing HMM and be recorded by the probe. To get the field profiles at the output plane with high resolution, the probe is mounted on an electric-controlled stage with scanning step of 1 mm. Each record data is averaged over 5 iterations to suppress the influence of noise. The computer is used to incorporate the whole process including controlling the stage by the controller box, reading data from the network analyzer and synthesizing all the operations.

We firstly experimental verify the directional propagation in the case of $\varepsilon_{e1} = 8.5$. Fig. 6(a) presents the averaged measured field intensity (blue solid line) and corresponding simulation results. The position of the source and scanning path are labeled by the dashed line in the simulated field maps (Fig. 6(b)), and the corresponding calculated structure is exactly the same as the experiment. We can notice the experiment is well matched with the simulations and two spikes are clearly distinguished. Therefore, we can conclude the EM waves generated by the source will propagate into the preferred directions inside the HMM. Then, we verify the ability of superresolution imaging by filling dielectric with $\varepsilon_{e2} = 4.7$ into the PPWG. Under the excited of a monopole antenna, the field will propagate forward without any tangential spread, as indicated by the simulated filed map in Fig. 6(d). The measured and simulated field intensities at the output face are plotted in Fig. 6(c). Remarkably, a resolution of 0.087λ is experimentally obtained (blue solid line), the value of which is much better than the previous HMM lens [13-17] owing to the low-loss. It is also recognized that the measured resolution is worse than the simulated results of 0.04λ. This can be attributed to the tolerances caused by fabrication, arrangement and measurement. By combing all the results, the effective-HMM is fairly verified, also demonstrating the ability of the ESSP in amplifying the evanescent waves.

## Discussion and Conclusion

Theoretically, the proposed methodology of amplifying the evanescent waves using common dielectrics is general at different wavelengths. The working mechanism is based on the structure-induced dispersion in a bounded waveguide, rather than the natural dispersion of the dielectrics. The permittivity can be freely tailored by the parameters of the waveguide, which has advantages over the conventional dielectric superlens in flexibility and low loss. In this technique, the conductivity of the metal should be large enough to support ideal waveguide modes, therefore exciting the ESSPs among dielectric layers. This requirement will limit the working frequency range of our method in the practical realization. To study the influences of the metal loss, dielectric loss and natural plasmonic effect when it comes to the terahertz and infrared ranges, we numerically investigate the imaging resolutions of effective HMM at 0.2 THz (λ = 1.5 mm), 2 THz (λ = 150 μm) and 20 THz (λ = 15 μm). SU-8, gallium arsenide (GaAs) and copper are set as the effective negative material, effective positive material and metal at terahertz wavelengths, i.e., 0.2 THz and 2 THz. The corresponding materials are replaced by calcium fluoride (CaF2), gallium phosphide (GaP) and Aluminum at far infrared (20 THz). The parameters of all the materials are taken from the experiment results and we also provide more choices for the effective HMM at different frequencies (more details about the simulations are put in section 4 of the supplement). The calculated field maps are illustrated in Figs. 7(a)-(c). It is found that, with the increase of working frequency, the loss issues become more dominant, leading to the degeneration of field intensity and imaging resolution. Besides, the plasmonic effects also have considerable influence of the TE1 mode, which enhances the value of effective permittivity. However, the designed structure still holds 0.085λ imaging resolution even at 20 THz. In Fig. 7(d), we plot the normalized intensity profiles at the imaging plane under the frequencies of 2.5 GHz, 0.2 THz, 2THz and 20 THz. A controlled sample without PPWG is calculated at 20 THz and its field map is put in the insert of Fig. 7(d). Although the working method is deteriorated by loss at 20 THz, the resolution is also enhanced 5 times when it is compared with the profile without PPWG, fully verifying the superlensing behavior induced by the ESSPs. Therefore, our methodology can break the conventional limitations of superresolution imaging from microwave to infrared wavelengths.

In conclusion, we propose and experimentally demonstrate that the broadband evanescent waves can be amplified by arbitrary positive dielectrics. The working principle is based on the effective plasmonic effects by exploiting the structural dispersion of TE1 mode in a PPWG. By using air as the effective plasmonic material, the spatial harmonics up to $7k_0$ is effectively enhanced and the superlens is also numerically verified. We further stack the air and dielectric layers to build the HMM. The validity of HMM is theoretically and experimentally demonstrated by the directional propagation and subwavelength resolution imaging with a measured resolution 0.087λ. We are confident that this work can provide more possibilities for beating the diffraction limit from microwave to infrared wavelengths. Besides, the proposed method is obviously not limited to this particular imaging demonstration. More applications based on the enhancing evanescent waves, such as ultrasensitive optics, spontaneous emission, light beam steering are also feasible.

## Acknowledgements

National Key Research and Development Program under Grant No. 2019YFA0210203;

National Natural Science Foundation of China under Grant No. 61971013.

**Conflict of Interest:** The authors declare no conflict of interest.

# Reference

(1) Zhang, X.; Liu, Z. Superlenses to overcome the diffraction limit. *Nat. Mater*. **2008**, 7, 435.

(2) Adam, A. J. L. Review of near-field terahertz measurement methods and their applications. *Infrared Millim. Terahertz Waves* **2011**, 32, 976.

(3) Rodriguez, R. D.; Madeira, T. I.; Sheremet, E.; Bortchagovsky, E.; Mukherjee, A.; Hietschold, M.; Zahn, D. R. Optical Absorption Imaging by Photothermal Expansion with 4 nm Resolution. *ACS Photonics* **2018**, 5, 3338-3346.

(4) Huang, T. J.; Tang, H. H.; Yin, L. Z.; Liu, J. Y.; Tan, Y.; Liu, P. K. Experimental demonstration of an ultra-broadband subwavelength resolution probe from microwave to terahertz regime. *Opt. Lett.* **2018**, 43, 3646-3649.

(5) Pendry, J. B. "Negative refraction makes a perfect lens," *Phys. Rev. Lett*. **2000**, 85, 3966–3969.

(6) Kawata, S.; Inouye, Y.; Verma, P. Plasmonics for near-field nano-imaging and superlensing. *Nature Photon.* **2009**, 3, 388.

(7) Fang, N.; Lee, H.; Sun, C.; Zhang, X. Sub–diffraction-limited optical imaging with a silver superlens. *Science* **2005**, 308, 534-537.

(8) Li, P.; Taubner, T. Broadband subwavelength imaging using a tunable graphene-lens. *ACS nano* **2012**, 6, 10107-10114.

(9) Adams, W.; Ghoshroy, A.; Güney, D. O. Plasmonic superlens imaging enhanced by incoherent active convolved illumination. *ACS Photonics* **2018**, 5, 1294-1302.

(10) Taubner, T.; Korobkin, D.; Urzhumov, Y.; Shvets, G.; Hillenbrand, R. Near-field microscopy through a SiC superlens. *Science* **2006**. 313, 1595-1595.

(11) Kehr, S. C.; McQuaid, R. G. P.; Ortmann. L.; Kämpfe, T.; Kuschewski, F.; Lang, D.; Eng, L. M. A local superlens. *ACS Photonics* **2015**, 3, 20-26.

(12) Fehrenbacher, M.; Winnerl, S.; Schneider, H.; Döring, J.; Kehr, S. C.; Eng, L. M.; Helm, M.; Plasmonic superlensing in doped GaAs. *Nano Lett*. **2015**, 15, 1057-1061.

(13) Lu, D; Liu, Z. Hyperlenses and metalenses for far-field super-resolution imaging. *Nat. Commun*. **2012**, 3, 1205.

(14) Liu, Z.; Lee, H.; Xiong, Y.; Sun, C.; Zhang, X. Far-field optical hyperlens magnifying sub-diffraction-limited objects. *Science* **2007**, 315, 1686-1686.

(15) Smolyaninov, I. I.; Hung, Y. J.; Davis, C. C. Magnifying superlens in the visible frequency range. *Science* **2007**, 5819, 1699-1701.

(16) Lee, D.; Kim, Y. D.; Kim, M.; So, S.; Choi, H. J.; Mun, J.; & Lee, H. Realization of wafer-scale hyperlens device for sub-diffractional biomolecular imaging. *ACS Photonics* **2017**, 5, 2549-2554.

(17) Rho, J.; Ye, Z., Xiong, Y.; Yin, X.; Liu, Z.; Choi, H.; Zhang, X. Spherical hyperlens for two-dimensional sub-diffractional imaging at visible frequencies. *Nat. Commun*. **2010**, 1, 143.

(18) Narimanov, E. Hyperstructured illumination. *ACS Photonics* **2016**, 3, 1090-1094.

(19) Poddubny, A.; Iorsh, I.; Belov, P.; Kivshar, Y. Hyperbolic metamaterials. *Nature Photon.* **2013**, 7, 948.

(20) Grigorenko, A. N.; Polini, M.; Novoselov, K. S. Graphene plasmonics. *Nature Photon.* **2012**, 6, 749.

(21) Shvets, G.; Trendafilov, S.; Pendry, J. B.; Sarychev, A. Guiding, focusing, and sensing on the subwavelength scale using metallic wire arrays. *Phys. Rev. Lett*. 99, 053903.

(22) Belov, P. A.; Zhao, Y.; Tse, S.; Ikonen, P.; Silveirinha, M. G.; Simovski, C. R.; Parini, C. Transmission of images with subwavelength resolution to distances of several wavelengths in the microwave range. *Phys. Rev. B* **2008**, 77, 193108.

(23) Huang, T. J.; Tang, H. H.; Tan, Y.; Li, L.; Liu, P. K. Terahertz Super-Resolution Imaging Based on Subwavelength Metallic Grating. *IEEE Trans. Antennas Propag*.

**2019**, 67, 3358-3365.

(24) Pendry, J. B.; Martin-Moreno, L.; Garcia-Vidal, F. J. Mimicking surface plasmons with structured surfaces. *Science* **2004**, 305, 847-848.

(25) Della Giovampaola, C.; Engheta, N. Plasmonics without negative dielectrics. *Phys. Rev. B* **2016**, 93, 195152.

(26) Ng, B.; Hanham, S. M.; Wu, J.; Fernández-Domínguez, A. I.; Klein, N.; Liew, Y. F.; Maier, S. A. Broadband terahertz sensing on spoof plasmon surfaces. *Acs Photonics* **2014**, 1, 1059-1067.

(27) Li, Z.; Liu, L.; Fernández-Domínguez, A. I.; Shi, J.; Gu, C.; García-Vidal, F. J.; Luo, Y. Mimicking Localized Surface Plasmons with Structural Dispersion. *Adv. Opt. Mater*. **2019**, 7, 1900118.

(28) Maier, S. A.; Andrews, S. R.; Martin-Moreno, L.; Garcia-Vidal, F. J. Terahertz surface plasmon-polariton propagation and focusing on periodically corrugated metal wires. *Phys. Rev. Lett*. **2006**, 97, 176805.

(29) Huang, T. J.; Liu, J. Y., Yin; L. Z.; Han, F. Y.; Liu, P. K. Superfocusing of terahertz wave through spoof surface plasmons. *Opt. Express* **2018**, 26, 22722-22732.

(30) Zhang, H. C.; Cui, T. J.; Zhang, Q.; Fan, Y.; Fu, X. Breaking the challenge of signal integrity using time-domain spoof surface plasmon polaritons. *ACS photonics* **2015**. 2, 1333-1340.

(31) Williams, C. R.; Andrews, S. R.; Maier, S. A.; Fernández-Domínguez, A. I.; Martín-Moreno, L.; García-Vidal, F. J. Highly confined guiding of terahertz surface plasmon polaritons on structured metal surfaces. *Nature Photon.* **2008**, 2, 175.

(32) Zhang, H. C.; Fan, Y.; Guo, J.; Fu, X.; Cui, T. J. Second-harmonic generation of spoof surface plasmon polaritons using nonlinear plasmonic metamaterials. *Acs Photonics* **2015**, 3, 139-146.

(33) Zhu, J. F.; Du, C. H.; Bao, L. Y.; Liu, P. K. Regenerated amplification of terahertz spoof surface plasmon radiation. *New J. Phys*. **2019**, 21, 033021.

(34) Yin, J. Y.; Ren, J.; Zhang, L.; Li, H.; Cui, T. J. Microwave Vortex-Beam Emitter Based on Spoof Surface Plasmon Polaritons. *Laser Photonics Rev.* **2018**, 12, 1600316.

(35) Su, H.; Shen, X.; Su, G.; Li, L.; Ding, J.; Liu, F.; Wang, Z. Efficient Generation of Microwave Plasmonic Vortices via a Single Deep-Subwavelength Meta-Particle. *Laser Photonics Rev.* **2018**, 12, 1800010.

(36) Tang, H. H.; Huang, B., Huang, T. J.; Tan, Y.; Liu, P. K. Efficient Waveguide Mode Conversions by Spoof Surface Plasmon Polaritons at Terahertz Frequencies. *IEEE Photonics J.* **2017**, 9, 1-10.

(37) Li, Z.; Liu, L.; Sun, H.; Sun, Y.; Gu, C.; Chen, X.; Luo, Y. Effective surface plasmon polaritons induced by modal dispersion in a waveguide. *Phys. Rev. Appl.* **2017**, 7, 044028.

(38) Li, Y.; Liberal, I.; Engheta, N. Structural dispersion–based reduction of loss in epsilon-near-zero and surface plasmon polariton waves. *Sci. Adv.* **2019**, 5(10), eaav3764.

(39) Yin, X.; Zhu, H.; Guo, H.; Deng, M.; Xu, T.; Gong, Z.; Chen, S. Hyperbolic metamaterial devices for wavefront manipulation. *Laser Photonics Rev.* **2019**, 13, 1800081.

(40) Huang, T. J.; Yin, L. Z.; Shuang, Y.; Liu, J. Y.; Tan, Y.; Liu, P. K. Far-field subwavelength resolution imaging by spatial spectrum sampling. *Phys. Rev. Appl.* **2019**, 12, 034046.

(41) Kehr, S. C.; Liu, Y. M.; Martin, L. W.; Yu, P.; Gajek, M.; Yang, S. Y.; Helm, M. Near-field examination of perovskite-based superlenses and superlens-enhanced probe-object coupling. *Nat. Commun.* **2011**, 2, 1-9.

(42) Noginov, M. A.; Gu, L.; Livenere, J.; Zhu, G.; Pradhan, A. K.; Mundle, R.; Podolskiy, V. A. Transparent conductive oxides: Plasmonic materials for telecom wavelengths. *Appl. Phys. Lett.* **2011**, 99, 021101.

(43) West, P. R.; Ishii, S.; Naik, G. V.; Emani, N. K.; Shalaev, V. M.; Boltasseva, A. Searching for better plasmonic materials. *Laser Photonics Rev.* **2010**, 4, 795-808.

(44) Naik, G. V.; Saha, B.; Liu, J.; Saber, S. M.; Stach, E. A.; Irudayaraj, J. M.; Boltasseva, A. Epitaxial superlattices with titanium nitride as a plasmonic component for optical hyperbolic metamaterials. *Proc. Natl. Acad. Sci.* **2014**, 111, 7546-7551.

(45) Boltasseva, A.; Atwater, H. A. Low-loss plasmonic metamaterials. *Science* **2011**, 331(6015), 290-291.

(46) Hoffman, A. J.; Alekseyev, L.; Howard, S. S.; Franz, K. J.; Wasserman, D.; Podolskiy, V. A.; Gmachl, C. Negative refraction in semiconductor metamaterials. *Nature Mater.* 2007, **6**, 946-950.

(47) Dai, S.; Ma, Q.; Andersen, T.; Mcleod, A. S.; Fei, Z.; Liu, M. K.; Keilmann, F. Subdiffractional focusing and guiding of polaritonic rays in a natural hyperbolic material. *Nat. Commun*. **2015**, 6, 1-7.

(48) Feng, K.; Harden, G.; Sivco, D. L.; Hoffman, A. J. Subdiffraction Confinement in All-Semiconductor Hyperbolic Metamaterial Resonators. *ACS Photonics* **2017**, 4, 1621-1626.

(49) Shen, L.; Wang, H.; Li, R.; Xu, Z.; Chen, H. Hyperbolic-polaritons-enabled dark-field lens for sensitive detection. *Sci. Rep.* **2017**, 7, 1-8.

(50) Huo, P.; Zhang, S.; Liang, Y.; Lu, Y.; Xu, T. Hyperbolic Metamaterials and Metasurfaces: Fundamentals and Applications. *Adv. Opt. Mater.* **2019**, 7, 1801616.

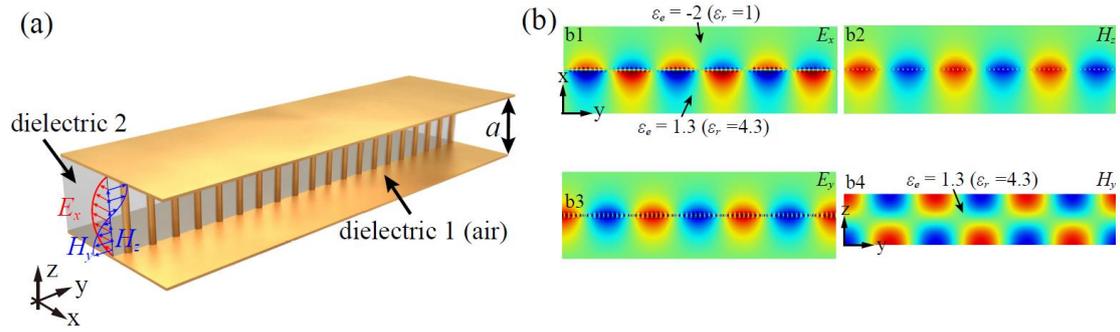

**Fig. 1.** (a) A PPWG is filled with two isotropic dielectric materials with the real permittivity being $\varepsilon_{r1}$ and $\varepsilon_{r2}$; (b) b1-b3 are the field distributions of $E_x$, $E_y$ and $H_z$ at the middle plane of the PPWG. b4 shows the snapshot of $H_y$ field at y-z plane, which is 0.4 mm away from the cylinder array.

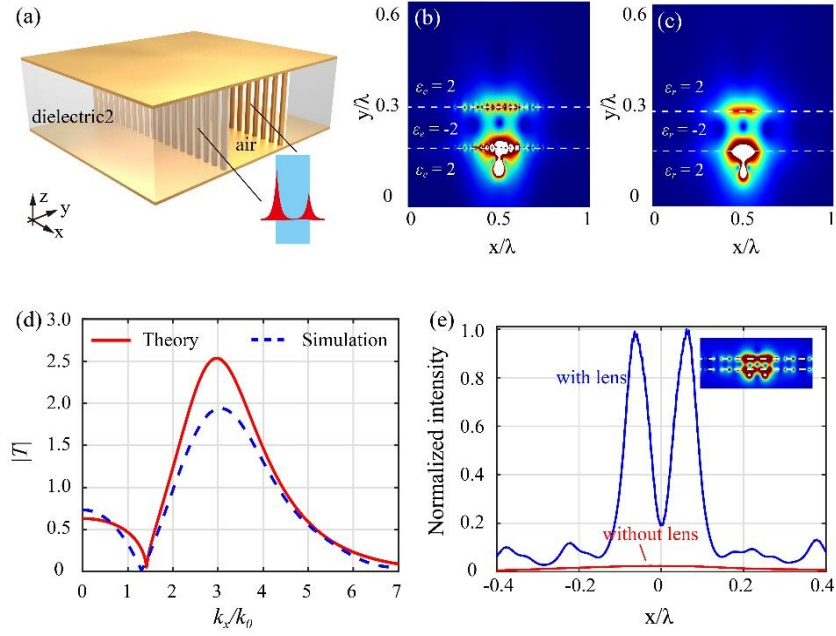

**Fig. 2.** (a) The schematic of the studied structure for amplifying evanescent waves. The inserted picture shows the process of amplification at two interfaces; (b) and (c) are electric field patterns of the effective and real plasmonic materials, respectively; (d) The theoretical (solid line) and simulated (dashed line) transfer functions of the air slab; (e) shows the line profiles of 0.1λ-separated two points taken from the imaging plane. In the insert of (e), we present the electric field distribution of the 'air' superlens.

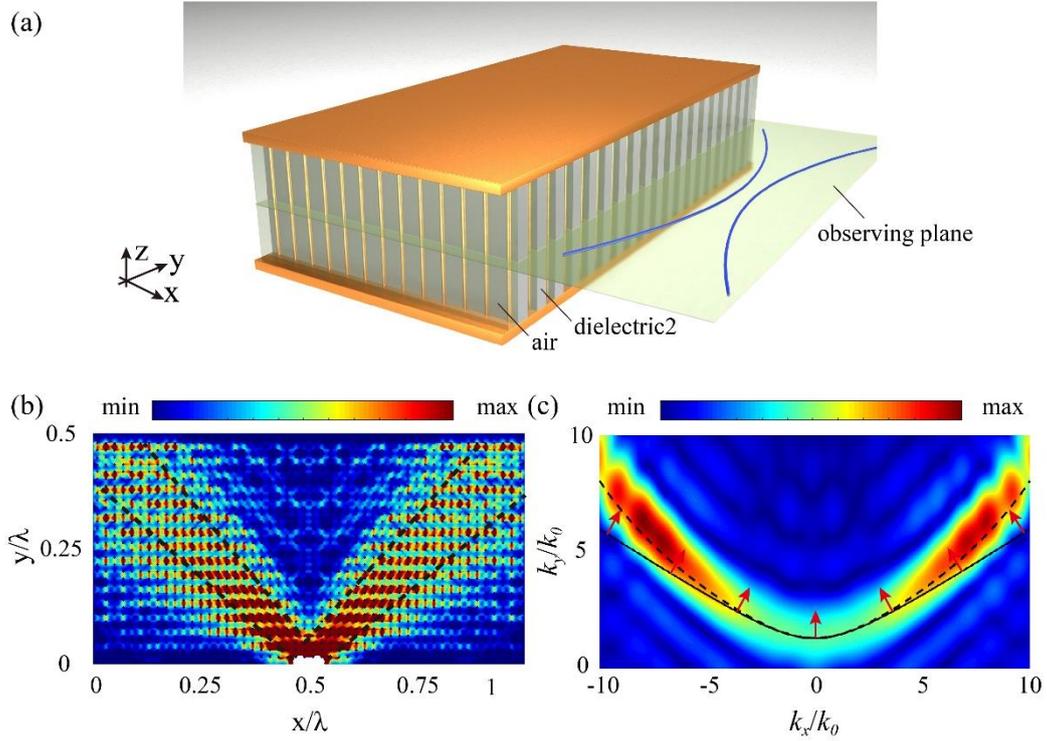

**Fig. 3.** (a) The designed structure of effective HMM, consisting of air and dielectric layers; (b) The simulated electric field distribution of HMM under the excitation of the slit.; (c) The dispersion of the designed HMM. The pseudo-color image is obtained by two-dimensional Fourier Transform of (b). The solid and dashed lines are the results calculated by effective medium theory without and with modification, respectively.

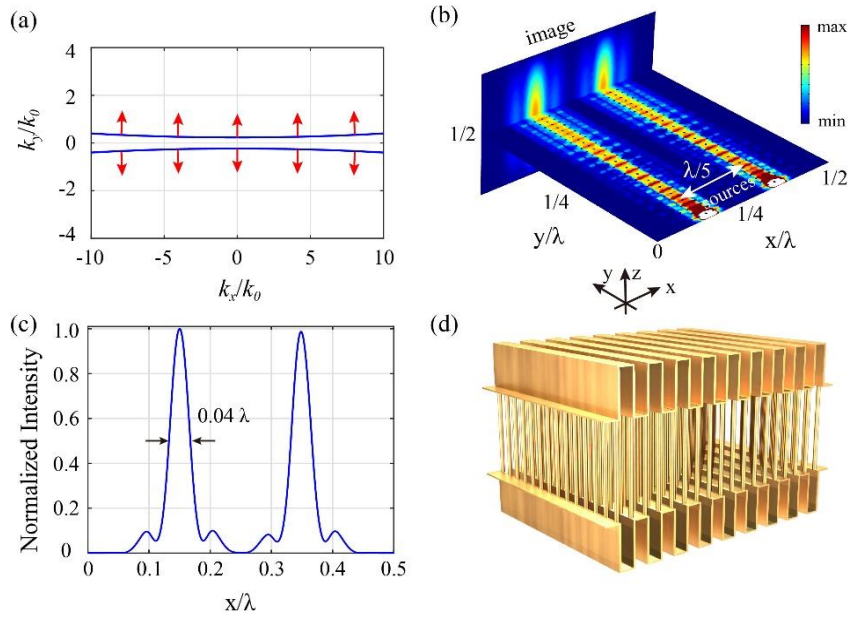

**Fig. 4.** (a) The dispersion curve of the design the HMM for the superresolution imaging; (b) illustrates the field snapshot of $H_z$ when two slits with $0.2\lambda$ separation are employed as the targets; (c) The profile of field intensity in the imaging plane of (b), indicating the ultimate resolution of $0.04\lambda$; (d) The magnetic field maps when the HMM is designed as separated shape.

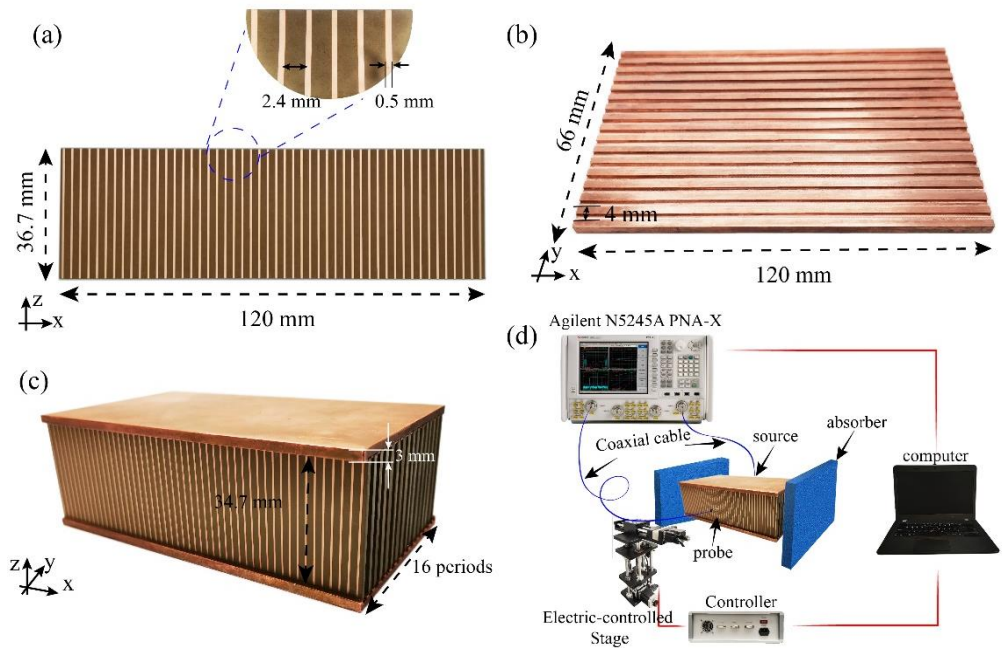

**Fig. 5.** Experimental realization. (a) The picture of fabricated dielectric layers, where the electrolytic copper-clad stripes are located on both sides. The inserted picture is the close shoot the stripes; (b) The copper plate with inner periodic grooves to mimic the PPWG and fix the dielectrics; (c) The assembled HMM structure by stack 16 periods of air and dielectric layers; (d) the schematic of experiment measurement.

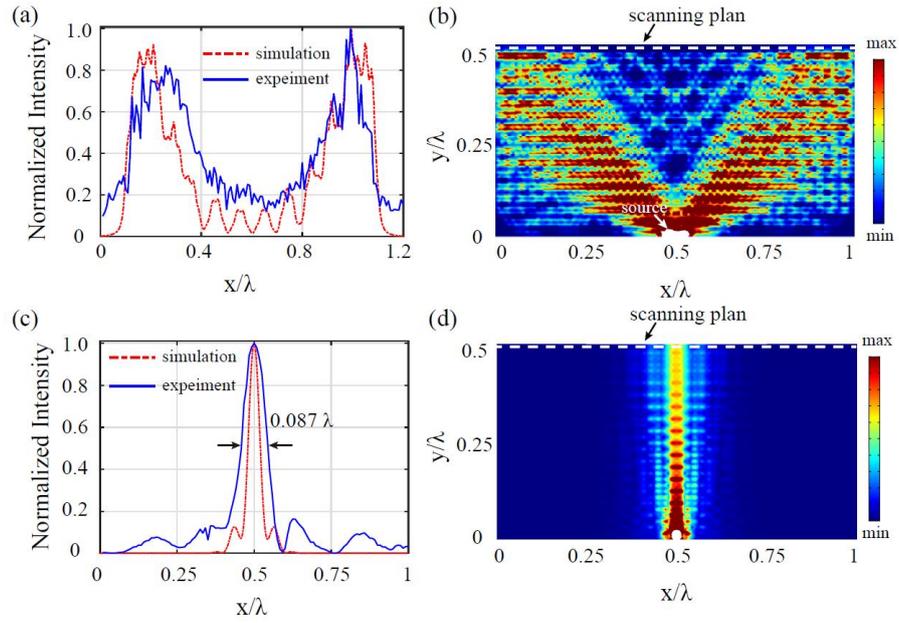

**Fig. 6.** (a) and (c) show the measured field intensity by scanning a monopole antenna, and the scanning lines are labeled in b) and b), the related simulated field distributions. In (a) and (b), the permittivity is set as 8.5 for the phenomenon of directional propagation; In (c) and (d), the permittivity is set as 4.7 for the subwavelength resolution imaging.

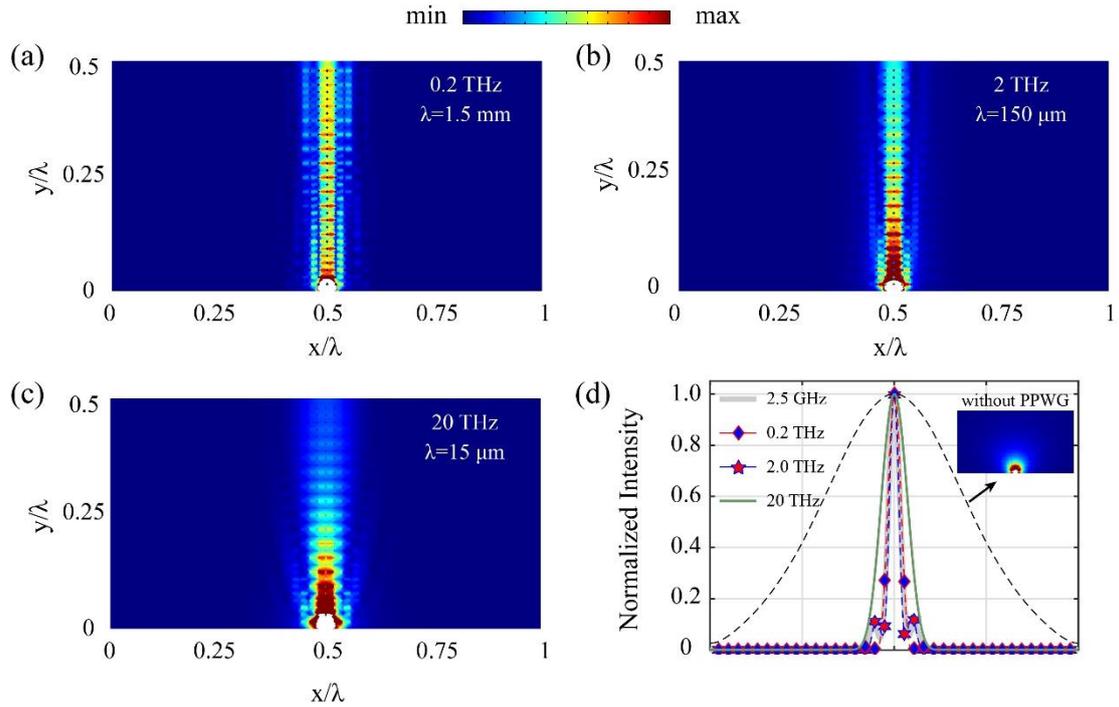

**Fig. 7.** The simulated results of HMM superlens from microwave to infrared wavelengths. (a)-(c) the calculated field distributions at 0.2 THz, 2 THz and 20 THz; (d) the intensity profiles at the imaging plane under serval calculated frequencies. The case of without PPWG is also calculated at 20 THz for comparison, and its field distribution is put in the insert.